\documentclass[12pt]{article}
\usepackage{graphicx}
\usepackage{latexsym}
\title{How (not) to teach Lorentz covariance of the Dirac equation}
\author{Hrvoje Nikoli\'c \\
Theoretical Physics Division, Rudjer Bo\v{s}kovi\'{c} Institute, \\
P.O.B. 180, HR-10002 Zagreb, Croatia \\
{\normalsize e-mail: hnikolic@irb.hr} \\
\makebox[1in]{} \\
}
\date{\today}
\begin{document}
\maketitle
\begin{abstract}
In the textbook proofs of Lorentz covariance of the Dirac equation,
one treats the wave function as a spinor and gamma matrices as scalars,
leading to a quite complicated formalism with several pedagogic drawbacks.  
As an alternative, I propose to teach Dirac equation and its Lorentz covariance
by using a much simpler, but physically equivalent formalism, in which 
these drawbacks do not appear. In this alternative formalism,
the wave function transforms as a scalar and gamma matrices as components
of a vector, such that the standard physically relevant bilinear combinations
do not change their transformation properties. 
The alternative formalism allows also a natural construction of some additional
non-standard bilinear combinations with well-defined transformation properties.  
\end{abstract}

\vspace{0.5cm}

\section{Introduction}

I like to ask tricky questions. For a warm up, here is a simple one appropriate to 
undergraduate students. Let $x$ be the position operator and $|\psi\rangle$ 
the quantum state. Which one of the two changes with time? No doubt, many students
will recall how these quantities appear in the Schr\"odinger equation, which will
lead them to the answer that $|\psi\rangle$ changes with time, while $x$ does not.
Certainly not a wrong answer, but there is a much better one: it depends on the {\em picture}. 
In the Schr\"odinger picture $|\psi\rangle$ changes with time and $x$ does not,
while in the Heisenberg picture $x$ changes with time and $|\psi\rangle$ does not.
This is consistent because neither $x$ nor $|\psi\rangle$ is a physical quantity by itself, 
while physical quantities do not depend on the picture.

Good! Now after this warm up, here is a tricky question that I really wanted to ask.
The question is appropriate to
graduate students, their teachers, and even experienced experts in
quantum field theory and particle physics. In the Dirac equation
\begin{equation}
 (i\gamma^{\mu}\partial_{\mu} -m)\psi =0 ,
\end{equation}
which of the two quantities, $\gamma^{\mu}$ and $\psi$, changes under a Lorentz transformation?
With very rare exceptions, almost everybody (assuming that they know what they are talking about) 
will answer that $\psi$ changes and $\gamma^{\mu}$
does not. Yet, quite analogously to the warm-up question above, that is not the best answer.
A much better answer is that it depends on the picture too. In the standard picture
known to everybody $\psi$ transforms and $\gamma^{\mu}$ does not,
but there is also an alternative picture in which $\gamma^{\mu}$ transforms and $\psi$ does not.
Neither $\gamma^{\mu}$ nor $\psi$ is a physical quantity by itself, while physical quantities 
do not depend on the picture.

Nevertheless, almost nobody will tell you about this alternative picture of $\gamma^{\mu}$ and $\psi$.
Why? Because that is not how they are taught. The purpose of the present paper is to
teach you that. And not only because the alternative picture exists, but also because it is
much simpler.

Fortunately, there is a relatively small community of physicists who are more likely to tell
you about the alternative picture, or at least recognize immediately that it exists
if you point it out to them.
These are people who work with spinors in {\em curved} spacetime. They know that 
the picture in which $\psi$ transforms and $\gamma^{\mu}$ doesn't is not appropriate 
for {\em general} coordinate transformations, of which Lorentz transformations are nothing but
a special case. Therefore, to deal with spinors in curved spacetime, they must first
``unlearn'' 
what they have learned about spinors in flat spacetime, and then learn again 
how to think about them in a different way. In this new picture,
it turns out \cite{birdel,GSW,parker} that $\gamma^{\mu}$ transforms as a vector
(which should not be surprising given the index ${\mu}$ it carries),
while $\psi$ does not transform because it is a scalar (which should 
also not be surprising given that it does not carry any vector index at all).

Unfortunately, the treatment of spinors in curved spacetime requires some advanced concepts
such as tetrads (called also {\em vierbeins}), with which people working in flat spacetime
are usually not familiar. Hence, it is not so simple to convey the idea 
to the flat-spacetime people by using the techniques developed by the curved-spacetime people.
Therefore, in this paper I develop a much simpler way to explain the alternative picture
of spinors in flat spacetime. 
(A simple remark that $\gamma^{\mu}$ and $\psi$ in flat spacetime may transform 
as a vector and a scalar, respectively, can also be found in \cite{hill}.)  
After the readers see this alternative picture, it is my hope that 
at least some of them will say: Wow, that's so simple,
why didn't they taught us spinors that way from the start?

The paper is organized as follows. In Sec.~\ref{SEC2}, I review the standard way of teaching
Lorentz covariance of the Dirac equation and discuss some pedagogical drawbacks of such teaching.
In Sec.~\ref{SEC3}, I formulate the alternative picture for the Dirac equation
which avoids these pedagogical drawbacks, and show 
that this alternative picture is much simpler and yet physically equivalent to the standard one.
This alternative picture can be taught even without 
ever referring to the standard one, as I outline in Sec.~\ref{SEC4}.
The conclusions are drawn in Sec.~\ref{SEC5}.

\section{Standard teaching and its drawbacks}  
\label{SEC2}

\subsection{Elements of standard teaching of Lorentz covariance of the Dirac equation}
\label{SEC2.1}

Let 
\begin{equation}\label{e1}
 (i\gamma^{\mu}\partial_{\mu} -m)\psi =0 
\end{equation}
be the Dirac equation in the Lorentz system ${\cal S}$, where $\gamma^{\mu}$ are the standard Dirac matrices
\cite{bd1} obeying
\begin{equation}\label{anticom}
 \{ \gamma^{\mu},\gamma^{\nu} \}=2\eta^{\mu\nu} ,
\end{equation}
and $\eta^{\mu\nu}={\rm diag}(1,-1,-1,-1)$ is the Minkowski metric.
Let 
\begin{equation}\label{e2}
 (i\gamma^{\mu}\partial'_{\mu} -m)\psi' =0 
\end{equation}
be the Dirac equation in another Lorentz system ${\cal S}'$.
(Here $\psi=\psi(x)$ and $\psi'=\psi'(x')$, but I do not write 
the $x$-dependence explicitly.) The Lorentz transformation of spacetime
coordinates can be written in the form
\begin{equation}\label{e3}
 x^{\mu}=\Lambda^{\mu}_{\;\nu}x'^{\nu} ,
\end{equation}
or more compactly in the matrix form $x=\Lambda x'$. The inverse of it is
$x'=\Lambda^{-1}x$, which in the component form reads
\begin{equation}\label{vector}
 x'^{\mu}=(\Lambda^{-1})^{\mu}_{\;\nu}x^{\nu} .
\end{equation}
Eq.~(\ref{e3}) implies
\begin{equation}\label{e4}
 \frac{\partial x^{\mu}}{\partial x'^{\nu}}=\Lambda^{\mu}_{\;\nu} .
\end{equation}
Since
\begin{equation}
\frac{\partial }{\partial x'^{\nu}}=\frac{\partial x^{\mu}}{\partial x'^{\nu}} \frac{\partial }{\partial x^{\mu}} ,
\end{equation}
(\ref{e4}) implies $\partial'_{\nu}=\Lambda^{\mu}_{\;\nu}\partial_{\mu}$, which we write as
\begin{equation}\label{e6}
\partial'_{\mu}=\Lambda^{\alpha}_{\;\mu}\partial_{\alpha} .  
\end{equation}
Therefore (\ref{e2}) can be written as
\begin{equation}\label{e7}
 (i\gamma^{\mu}\Lambda^{\alpha}_{\;\mu}\partial_{\alpha} -m)\psi' =0 . 
\end{equation}

Next write
\begin{equation}\label{e8}
 \psi'=S\psi ,
\end{equation}
where $S$ is some $x$-independent matrix the properties of which need to be determined.
For that purpose one multiplies (\ref{e1}) (with $\mu\rightarrow\alpha$) 
with $S$ from the left and inserts $1=S^{-1}S$, so
\begin{equation}\label{e9}
 iS\gamma^{\alpha}S^{-1}\partial_{\alpha}S\psi -mS\psi =0 . 
\end{equation}
Using (\ref{e8}), this can be written as
\begin{equation}\label{e10}
 (iS\gamma^{\alpha}S^{-1}\partial_{\alpha} -m)\psi' =0 . 
\end{equation}
Comparing (\ref{e10}) with (\ref{e7}), one obtains
\begin{equation}\label{e11}
 S\gamma^{\alpha}S^{-1}=\gamma^{\mu}\Lambda^{\alpha}_{\;\mu} .
\end{equation}
From this equation one can determine $S$ as a function of $\Lambda^{\mu}_{\;\nu}$, but the procedure
is quite complicated (see e.g. \cite{bd1}), so I omit it. But even without the explicit expression
for $S$ as a function of $\Lambda^{\mu}_{\;\nu}$, it is clear that the inverse Lorentz transformation
must correspond to the inverse $S$. Therefore (\ref{e11})
can also be written as
\begin{equation}\label{e11'}
 S^{-1}\gamma^{\alpha}S=\gamma^{\mu}(\Lambda^{-1})^{\alpha}_{\;\mu} .
\end{equation}

Using (\ref{e11}), 
it is possible to prove that
\begin{equation}\label{inverse}
S^{-1}=\gamma^0 S^{\dagger} \gamma^0.
\end{equation}
(Unfortunately, I am not aware of any simple proof of (\ref{inverse}). 
The simplest proof I am aware, but still quite involved,
is presented in \cite{schweber}.)
By multiplying (\ref{inverse}) with $\gamma^0$ from the left and using $\gamma^0\gamma^0=1$
(which follows directly from (\ref{anticom})),
one gets a very useful form of (\ref{inverse})
\begin{equation}\label{inverse2}
\gamma^0 S^{-1}=S^{\dagger} \gamma^0.
\end{equation}

An important consequence of (\ref{inverse}) is that $S^{-1} \neq S^{\dagger}$, i.e. $S^{\dagger}S\neq 1$.
Therefore
\begin{equation}\label{psidpsi}
 (\psi^{\dagger}\psi)'=\psi^{\dagger}S^{\dagger}S \psi \neq \psi^{\dagger} \psi ,
\end{equation}
which shows that $\psi^{\dagger}\psi$ does not transform as a scalar. 
On the other hand, defining
\begin{equation}
 \bar{\psi}=\psi^{\dagger} \gamma^0 
\end{equation}
and using (\ref{inverse2}) one obtains
\begin{equation}
 (\bar{\psi}\psi)'=\psi^{\dagger}S^{\dagger}\gamma^0 S \psi = \psi^{\dagger}\gamma^0 S^{-1}S \psi 
= \psi^{\dagger}\gamma^0 \psi = \bar{\psi}\psi ,
\end{equation}
which shows that $\bar{\psi}\psi=\psi^{\dagger}\gamma^0\psi$ transforms as a scalar.

In a similar way one finds
\begin{equation}\label{dircur}
 (\bar{\psi}\gamma^{\mu}\psi)'=\psi^{\dagger}S^{\dagger}\gamma^0\gamma^{\mu}S \psi =
\psi^{\dagger}\gamma^0 S^{-1}\gamma^{\mu}S \psi .
\end{equation}
Using (\ref{e11'}), it can be written as
\begin{equation}\label{dircur2}
 (\bar{\psi}\gamma^{\mu}\psi)'= (\Lambda^{-1})^{\mu}_{\;\nu} \; \psi^{\dagger}\gamma^0\gamma^{\nu}\psi 
=(\Lambda^{-1})^{\mu}_{\;\nu} \; \bar{\psi}\gamma^{\nu}\psi ,
\end{equation}
so comparing it with (\ref{vector}) one concludes that 
$\bar{\psi}\gamma^{\mu}\psi$ transforms as a vector.

\subsection{The drawbacks of standard teaching}

From Sec.~\ref{SEC2.1} one can see that the Lorentz covariance of the Dirac equation
is quite complicated. For comparison, Lorentz covariance of the Maxwell equations is much simpler.
If possible, it would certainly be desirable to have a simpler formulation of the
Lorentz covariance for the Dirac equation. 

Moreover, there is something potentially confusing about the standard teaching
outlined in Sec.~\ref{SEC2.1}. The notation $\gamma^{\mu}$ suggests that this object also might
transform as a vector. Why then $\gamma^{\mu}$ in (\ref{e2}) is not replaced by $\gamma'^{\mu}$?
Most textbooks which discuss Lorentz covariance of the Dirac equation, 
including those by Schweber \cite{schweber}, Sakurai \cite{sakurai},
Itzykson and Zuber \cite{zuber}, and Zee \cite{zee}, do not attempt to answer that question.
From the pedagogical point of view, this is certainly not the best way to teach 
Lorentz covariance of the Dirac equation.

In some textbooks, including those by Bjorken and Drell \cite{bd1}, Messiah \cite{messiah},
Jauch and Rohrlich \cite{jauch}, and Greiner \cite{greiner},
a somewhat better approach is exploited.
They note that $\gamma'^{\mu}$ is not equal to $\gamma^{\mu}$, 
but explain that they are related by a unitary transformation.
Consequently, their argument goes, the $\gamma^{\mu}$ can be fixed
and viewed as objects that do not transform under Lorentz transformations.
Nevertheless, from the pedagogical point of view, such an approach is also
not completely satisfying. A unitary transformation which transforms
$\gamma'^{\mu}$ back to $\gamma^{\mu}$ should affect also the spinor $\psi$. 
So, how would $\psi$ transform if one did {\em not} choose to transform 
$\gamma'^{\mu}$ back to $\gamma^{\mu}$? In that case, would $\psi$ still transform as 
a spinor? Or would it perhaps become a scalar?
The textbooks above say nothing about that, so it is also not the perfect way
to teach Lorentz covariance of the Dirac equation. 

The two approaches above have in common that they first introduce the Dirac equation, and then
show that $\psi$ transforms in a specific way, known as transformation of spinors.
As an alternative, some textbooks, including those by ``Landau and Lifshitz'' \cite{landau},
Ryder \cite{ryder}, and Weinberg \cite{weinberg},  
choose a reversed pedagogy. They first introduce the concept of spinors as abstract algebraic
objects (that even do not need to depend on $x$), and then introduce the Dirac equation 
as an application of spinor mathematics to physics. No doubt, such an approach offers a much 
deeper mathematical understanding of spinors. In particular, their transformation properties
are obtained without referring to the Dirac equation. In addition, the mathematical
origin of $\gamma^{\mu}$ matrices is explained, from which it becomes clear why they are
fixed matrices which do not transform. Nevertheless, even that mathematically more 
sophisticated approach is not perfect from the pedagogical point of view, 
precisely because it is mathematically sophisticated. Namely, the mathematical sophistication
makes the theory even more complicated, which many practically oriented physicists
view as an unnecessary distraction from their true goal -- learning {\em physics}.
 
Finally, there are many textbooks which study Dirac equation but do not really attempt to prove its 
Lorentz covariance. Such books may be excellent for teaching what they really want to 
teach (e.g. how to calculate the scattering amplitude for elementary particles),
but in the context of teaching Lorentz covariance of the Dirac equation
they do not deserve to be mentioned.

Let me end this section with an exercise for the reader. Take three books which study the Dirac equation, 
not all of which are mentioned in the list of references for this paper. 
For each of the three books, answer the following questions: Are spinors introduced before or after
introducing the Dirac equation? Is Lorentz covariance of the Dirac equation proved?
Is it explicitly stated that $\gamma^{\mu}$ does not transform under Lorentz transformations?
If yes, is it explained why? Is it written down explicitly how $\psi$ transforms
under Lorentz transformations? If yes, is that transformation law derived?

\section{Two pictures for the Dirac equation}
\label{SEC3}

In Sec.~\ref{SEC2}, I have studied the standard picture for the Dirac equation, in which
the wave function $\psi$ transforms as a spinor under Lorentz transformations of spacetime coordinates, 
while the gamma matrices $\gamma^{\mu}$ do not transform at all. Since the transforming quantity transforms
as a spinor, I refer to this picture as {\em spinor} picture.

Here I introduce a different picture for the Dirac equation, in which 
the wave function does not transform under Lorentz transformations of spacetime coordinates,
while the gamma matrices transform as components of a vector.
Since the transforming quantity transforms
as a vector, I refer to this picture as {\em vector} picture.
To distinguish the wave function and gamma matrices in the vector picture 
from those in the spinor picture, those in the vector picture are denoted by
$\Psi$ and $\Gamma^{\mu}$ respectively.

The Dirac equation (\ref{e1}) in the vector picture reads
\begin{equation}\label{e12}
 (i\Gamma^{\mu}\partial_{\mu} -m)\Psi =0 . 
\end{equation}
Since I postulate that $\Gamma^{\mu}$ transforms as a vector, (\ref{e3}) implies that it transforms according to 
\begin{equation}\label{e13}
\Gamma^{\mu}=\Lambda^{\mu}_{\;\nu} \Gamma'^{\nu} .
\end{equation}
Likewise, postulating that $\Psi$ is a scalar means 
\begin{equation}\label{e14}
 \Psi'=\Psi .
\end{equation} 

Since $\Gamma^{\mu}$ is a vector and $\Psi$ is a scalar, the Lorentz covariance
of (\ref{e12}) is 
quite trivial. Nevertheless, for the sake of completeness, let me present the proof explicitly.
Eq.~(\ref{e13}) can be inverted as 
\begin{equation}\label{e13inv}
\Gamma'^{\mu}=(\Lambda^{-1})^{\mu}_{\;\nu} \Gamma^{\nu} .
\end{equation} 
This together with (\ref{e14}) and (\ref{e6}) gives
\begin{eqnarray}\label{proof_covar}
\Gamma'^{\mu}\partial'_{\mu} \Psi' 
&=& (\Lambda^{-1})^{\mu}_{\;\nu} \Gamma^{\nu} \Lambda^{\alpha}_{\;\mu}\partial_{\alpha} \Psi
= \Lambda^{\alpha}_{\;\mu} (\Lambda^{-1})^{\mu}_{\;\nu} \Gamma^{\nu} \partial_{\alpha} \Psi
\nonumber \\
&=& (\Lambda \Lambda^{-1})^{\alpha}_{\;\nu} \Gamma^{\nu} \partial_{\alpha} \Psi =
\delta^{\alpha}_{\nu} \Gamma^{\nu} \partial_{\alpha} \Psi =  \Gamma^{\alpha}\partial_{\alpha} \Psi.
\end{eqnarray}
This means that $\Gamma'^{\mu}\partial'_{\mu} \Psi'=\Gamma^{\mu}\partial_{\mu} \Psi$,
which shows that (\ref{e12}) is Lorentz covariant.
%
%
Note that this simple proof does not depend on Eq.~(\ref{e11}).  

The non-trivial aspects of (\ref{e12}), however, are (i) to find out how $\Psi$ and $\Gamma^{\mu}$ 
are related to $\psi$ and $\gamma^{\mu}$, and (ii) to prove that (\ref{e12}) is equivalent to
(\ref{e1}). 
This is what I do next.
(In particular, unlike the proof of Lorentz covariance in (\ref{proof_covar}),
the proof of equivalence of the two pictures will depend on (\ref{e11}).)

As the transformation properties of $\Psi$, $\Gamma^{\mu}$, $\psi$ and $\gamma^{\mu}$ are defined, 
to establish the general relation between them it is sufficient to specify the relation in one
particular Lorentz system of coordinates. For convenience it can be chosen to be the laboratory system
${\cal S}_{\rm lab}$, in which I choose
\begin{equation}\label{e15}
 \Gamma^{\mu}_{\rm lab}=\gamma^{\mu} ,\;\;\; \Psi=\psi_{\rm lab} .
\end{equation}
In this sense the laboratory system can be thought of as a ``preferred'' system of coordinates,
but it does not ruin the Lorentz covariance of the vector picture, because the 
only purpose of the ``preferred'' system is to establish the relation between the two pictures.
Indeed, to use the analogy from the Introduction, 
this is very much analogous to the fact that the operators and states in the Heisenberg picture coincide
with those in the Schr\"odinger picture at one particular ``initial'' value of time $t_0$, but it does
not ruin the fact that each of the pictures by itself is invariant under time translations.
Moreover, I show in the Appendix how $\Psi$ and $\Gamma^{\mu}$
can be defined in a mathematically more elegant way,
without explicitly referring to any particular system of coordinates. 
 
Now, the rest of analysis is straightforward. Eq.~(\ref{e12}) in the system ${\cal S}_{\rm lab}$ is
\begin{equation}\label{e16}
 (i\Gamma^{\mu}_{\rm lab}\partial_{\mu}^{\rm lab} -m)\Psi =0 . 
\end{equation}
Using (\ref{e15}), this can be written as
\begin{equation}\label{e19}
 (i\gamma^{\mu}\partial_{\mu}^{\rm lab} -m)\psi_{\rm lab} =0 . 
\end{equation}
Now take $\Lambda$ to be the Lorentz transformation that connects the system ${\cal S}$ with the system
${\cal S}_{\rm lab}$, so that (\ref{e8}) and (\ref{e6}) become
\begin{equation}\label{e20}
 \psi_{\rm lab}=S\psi ,
\end{equation}
\begin{equation}\label{e21}
\partial^{\rm lab}_{\mu}=\Lambda^{\alpha}_{\;\mu}\partial_{\alpha} .  
\end{equation}
In this way (\ref{e19}) can be written as
\begin{equation}\label{e22}
 (i\gamma^{\mu} \Lambda^{\alpha}_{\;\mu} \partial_{\alpha} -m)S\psi =0 , 
\end{equation}
which after using (\ref{e11}) becomes
\begin{equation}\label{e23}
 iS\gamma^{\alpha}S^{-1}S \partial_{\alpha}\psi -mS\psi =0 . 
\end{equation}
Hence, by multiplying with $S^{-1}$ from the left one finally obtains
\begin{equation}\label{e25}
 (i\gamma^{\alpha} \partial_{\alpha} -m)\psi =0 . 
\end{equation}
In this way, from the Dirac equation in the vector picture (\ref{e12}) I have derived
the Dirac equation in the spinor picture (\ref{e25}). The derivation can also be inverted
step by step, implying that starting from the Dirac equation in the spinor picture (\ref{e25})
one can derive the Dirac equation in the vector picture (\ref{e12}).
This proves that {\em the two pictures are equivalent}.

It is also instructive to see how some bilinear combinations of $\Psi$ are related to those
of $\psi$. I first define 
\begin{equation}
 \bar{\Psi}=\Psi^{\dagger} \gamma^0 , 
\end{equation}
which is clearly a scalar. Therefore, it is obvious that $\bar{\Psi}\Psi$ is a scalar and
$\bar{\Psi}\Gamma^{\mu}\Psi$ a vector, so it does not need to be proved.
What needs to be proved is that they are equal to $\bar{\psi}\psi$ and $\bar{\psi}\gamma^{\mu}\psi$,
respectively.
For $\bar{\Psi}\Psi$, I perform a straightforward proof
\begin{equation}\label{scalarbar}
 \bar{\Psi}\Psi=\Psi^{\dagger}\gamma^0\Psi=\psi^{\dagger}_{\rm lab}\gamma^0\psi_{\rm lab}=
\bar{\psi}_{\rm lab}\psi_{\rm lab}=(\bar{\psi}\psi)_{\rm lab}=\bar{\psi}\psi ,
\end{equation}
where in the last equality I have used the fact that $\bar{\psi}\psi$ is a scalar.
For $\bar{\Psi}\Gamma^{\mu}\Psi$ the simplest way is to use use a trick. 
Since it is known that both $\bar{\Psi}\Gamma^{\mu}\Psi$
and $\bar{\psi}\gamma^{\mu}\psi$ transform as vectors, it is sufficient to show that they are equal
in one particular Lorentz system. But they are obviously equal in the laboratory system,
because in that system $\bar{\Psi}=\bar{\psi}$, $\Gamma^{\mu}=\gamma^{\mu}$, and $\Psi=\psi$.
Therefore $\bar{\Psi}\Gamma^{\mu}\Psi$ is equal to $\bar{\psi}\gamma^{\mu}\psi$ in all
Lorentz systems, which finishes the proof.

I have shown above that
\begin{equation}
 \bar{\Psi}\Psi=\bar{\psi}\psi , \;\;\; \bar{\Psi}\Gamma^{\mu}\Psi=\bar{\psi}\gamma^{\mu}\psi .
\end{equation}
The same can be shown for other similar bilinear combinations of $\psi$ . 
Since physical quantities are expressed in terms of such bilinear combinations,
it shows that the two pictures for the Dirac equation are {\em physically equivalent}.

The advantage of the vector picture is that the proof of its Lorentz covariance is much simpler.
However, that is not the only advantage. The vector picture allows a simple and natural construction
of some additional bilinear combinations with well-defined transformation properties,
which in the spinor picture cannot be constructed so naturally.
The two most interesting combinations are
\begin{equation}\label{rho}
 \rho = \Psi^{\dagger}\Psi , 
\end{equation}
\begin{equation}\label{KG}
j_{\mu} = \frac{i}{2} \, \Psi^{\dagger} \stackrel{\leftrightarrow\;}{\partial_{\mu}} \Psi ,
\end{equation}
where $A\!\stackrel{\leftrightarrow\;}{\partial_{\mu}}\!B \equiv A (\partial_{\mu} B)-(\partial_{\mu} A) B$.
Clearly, (\ref{rho}) transforms as a scalar and (\ref{KG}) transforms as a vector.
Their possible physical interpretation is discussed in \cite{nikIJMPA10,nikBOOK12,nik_timeprob}. 

For someone who never heard about the vector picture (which refers to the large majority of physicists
at the time of writing this paper), it may be hard to believe that
(\ref{rho}) and (\ref{KG}) transform as a scalar and a vector, respectively. In particular,
isn't the claim that (\ref{rho}) transforms as a scalar in contradiction with the fact that 
(\ref{psidpsi}) does not transform as a scalar? Let me show that there is no contradiction,
by attempting to find a contradiction and seeing how exactly the attempt fails. Similarly to 
(\ref{scalarbar}), one obtains
\begin{equation}\label{e28}
 \Psi^{\dagger}\Psi=\Psi^{\dagger}\gamma^0\gamma^0\Psi=\psi^{\dagger}_{\rm lab}\gamma^0\gamma^0\psi_{\rm lab}=
\bar{\psi}_{\rm lab}\gamma^0\psi_{\rm lab} .
\end{equation}
Naively one might think that the last quantity $\bar{\psi}_{\rm lab}\gamma^0\psi_{\rm lab}$
transforms as a time-component of a vector, which would contradict the claim
that $\Psi^{\dagger}\Psi$ transforms as a scalar. But in fact $\bar{\psi}_{\rm lab}\gamma^0\psi_{\rm lab}$
does {\em not} transform as a time-component of a vector, because
\begin{equation}\label{e29}
 \bar{\psi}_{\rm lab}\gamma^0\psi_{\rm lab} \neq \bar{\psi}\gamma^0\psi .
\end{equation}
The two quantities in
(\ref{e29}) are equal only if $\psi$ is evaluated in the laboratory system, but in general
they are different.
Therefore there is no contradiction between the facts that $\Psi^{\dagger}\Psi$ and 
$\psi^{\dagger}\psi=\bar{\psi}\gamma^0\psi$
transform as a scalar and a time-component of a vector, respectively. These two quantities
coincide in one Lorentz system (chosen to be the laboratory one), but in other Lorentz systems
they are different quantities. 
A more formal demonstration of this fact is given also in the Appendix.

\section{Teaching only the vector picture}
\label{SEC4}

We have seen that there are two equivalent pictures for the Dirac equation: the standard
spinor picture and the alternative vector picture. We have also seen that the alternative
vector picture is much simpler. Therefore, in this section I propose an alternative way to teach
the Dirac equation, by teaching {\em only} the vector picture. Of course, a drawback
of such teaching would be a clash with most of the existing literature, which could create confusion.
Nevertheless, given the advantages of such teaching, I believe it is worthwhile at least
to outline how such teaching might look like. So this is what I do in what follows.

In an attempt to linearize the Klein-Gordon equation
\begin{equation}\label{eKG}
 (\partial^{\mu}\partial_{\mu} +m^2)\Psi=0 ,
\end{equation}
one obtains the Dirac equation
\begin{equation}\label{e12v}
 (i\Gamma^{\mu}\partial_{\mu} -m)\Psi =0 , 
\end{equation}
where
\begin{equation}\label{anticomv}
 \{ \Gamma^{\mu},\Gamma^{\nu} \}=2\eta^{\mu\nu} .
\end{equation}
Here $\Gamma^{\mu}$ transforms as a vector and $\Psi$ as a scalar under Lorentz transformations of 
spacetime coordinates, so the Lorentz covariance of (\ref{e12v}) is obvious.
However, (\ref{anticomv}) suggests that $\Gamma^{\mu}$ should be $n\times n$ matrices, so $\Psi$
should be an $n$-component column. It turns out that the smallest possible value of $n$ is 4,
so one fixes $n=4$. One special choice for $\Gamma^{\mu}$ satisfying (\ref{anticomv})
are the standard Dirac matrices $\gamma^{\bar{\mu}}$.
(In the rest of the paper they are denoted by $\gamma^{\mu}$, but here I modify the notation by 
putting the bar over $\mu$ which reminds us that $\bar{\mu}$ is not a vector index.
Namely, the $\gamma^{\bar{\mu}}$ are fixed matrices which do not transform under Lorentz transformations,
which is why the notation $\gamma^{\bar{\mu}}$ is better than $\gamma^{\mu}$.) 
Thus one may determine the vector $\Gamma^{\mu}$ 
in any Lorentz system by choosing one particular Lorentz system, say the laboratory one, in which
\begin{equation}
 \Gamma^{\mu}_{\rm lab}=\gamma^{\bar{\mu}} .
\end{equation}

Defining
\begin{equation}
 \bar{\Psi}=\Psi\gamma^{\bar{0}} ,
\end{equation}
the Dirac equation (\ref{e12v}) implies that the Dirac vector current
\begin{equation}
 j^{\mu}_{\rm Dirac}=\bar{\Psi}\Gamma^{\mu}\Psi
\end{equation}
is conserved
\begin{equation}
 \partial_{\mu}j^{\mu}_{\rm Dirac}=0.
\end{equation}
Similarly, the Klein-Gordon equation (\ref{eKG}) implies that the Klein-Gordon vector current
\begin{equation}\label{KGv}
j_{\mu} = \frac{i}{2} \, \Psi^{\dagger} \stackrel{\leftrightarrow\;}{\partial_{\mu}} \Psi 
\end{equation}
is conserved too, i.e.
\begin{equation}
 \partial_{\mu}j^{\mu}=0.
\end{equation}
In most physical applications only the Dirac current is relevant, but in some 
applications the Klein-Gordon current may be relevant as well.
Similarly, one can construct two bilinear scalars $\bar{\Psi}\Psi$ and $\Psi^{\dagger}\Psi$.
In most physical applications only $\bar{\Psi}\Psi$ is relevant,
but in some applications $\Psi^{\dagger}\Psi$ may be of interest as well.

\section{Conclusion}
\label{SEC5}

In this paper, I have identified some pedagogical drawbacks in the standard approaches
to teaching Lorentz covariance of the Dirac equation, including the fact that the proof
of Lorentz covariance is quite complicated. To avoid these drawbacks, I have proposed
an alternative way to teach Lorentz covariance of the Dirac equation, by 
introducing a new formalism.  
The proposed formalism is inspired by the treatment of spinors in curved spacetime,
but is in fact much simpler than that (because it is formulated in flat spacetime)
and logically independent of it. The main idea of the formalism is to perform
a transformation from the standard $\psi$ and $\gamma^{\mu}$, which transform as a spinor
and a scalar, respectively, to new quantities $\Psi$ and $\Gamma^{\mu}$, which transform as  
a scalar and a vector, respectively. I have shown that the two formalisms are physically 
equivalent, but that the new formalism is much simpler and can be taught even without
referring to the standard formalism.
In addition, the new formalism allows a natural construction of some non-standard 
bilinear combinations with well-defined transformation properties, such as the
vector Klein-Gordon current and the scalar $\Psi^{\dagger}\Psi$.

\section*{Acknowledgments}

The author is grateful to B. Klajn and B. Ni\v{z}i\'c for useful and encouraging comments on the manuscript,
and to S. Dowker for drawing attention to some relevant references. 
This work was supported by the Ministry of Science of the
Republic of Croatia under Contract No.~098-0982930-2864.

\appendix

\section{Formal transformation theory between spinor and vector picture of the Dirac equation}

By starting from the standard spinor picture described in Sec.~\ref{SEC2.1}, 
in this section I rederive the main results of Sec.~\ref{SEC3} by
developing the formal transformation theory between spinor and vector
picture without explicitly referring to any particular system of coordinates.

Starting from (\ref{e8}), consider the transformation
\begin{equation}
 (S^{-1}\psi)'=SS^{-1}\psi=\psi .
\end{equation}
This shows that $S^{-1}\psi$ transforms as a scalar, so I define the scalar
\begin{equation}\label{app2}
 \Psi=S^{-1}\psi .
\end{equation}
Since $\Psi$ is a scalar, the quantity 
\begin{equation}\label{app3}
 \bar{\Psi}=\Psi^{\dagger} \gamma^0 
\end{equation} 
is also a scalar. Using (\ref{app2}), (\ref{app3}) can also be written as
\begin{equation}\label{app4}
 \bar{\Psi}=\psi^{\dagger} (S^{-1})^{\dagger} \gamma^0.
\end{equation}
I want to define a quantity $\Gamma^{\mu}$ by requiring that
\begin{equation}\label{app5}
 \bar{\Psi}\Gamma^{\mu}\Psi = \bar{\psi}\gamma^{\mu}\psi .
\end{equation}
The right-hand side of (\ref{app5}) is a vector while on the left-hand side
$\bar{\Psi}$ and $\Psi$ are scalars, which implies that $\Gamma^{\mu}$ is a vector.
What I need is a relation between $\Gamma^{\mu}$ and $\gamma^{\mu}$.
For that purpose, I write
\begin{eqnarray}\label{app6}
\bar{\Psi}\Gamma^{\mu}\Psi & = & \psi^{\dagger} (S^{-1})^{\dagger} \gamma^0 \Gamma^{\mu} S^{-1}\psi
\nonumber \\
& = & \psi^{\dagger} (S^{-1})^{\dagger} \gamma^0 S^{-1}S \Gamma^{\mu} S^{-1}\psi
\nonumber \\
& = & \psi^{\dagger} (S^{-1})^{\dagger} S^{\dagger} \gamma^0 S \Gamma^{\mu} S^{-1}\psi
\nonumber \\
& = & \psi^{\dagger} (SS^{-1})^{\dagger} \gamma^0 S \Gamma^{\mu} S^{-1}\psi
\nonumber \\
& = & \psi^{\dagger} \gamma^0 S \Gamma^{\mu} S^{-1}\psi = \bar{\psi} S \Gamma^{\mu} S^{-1}\psi ,
\end{eqnarray}
where in the third line I used (\ref{inverse2}).
The comparison with (\ref{app5}) shows that 
\begin{equation}
\gamma^{\mu}=S \Gamma^{\mu} S^{-1} ,
\end{equation}
which is equivalent to
\begin{equation}\label{app7}
 \Gamma^{\mu} = S^{-1} \gamma^{\mu} S .
\end{equation}

Now multiply the Dirac equation $(i\gamma^{\mu}\partial_{\mu} -m)\psi =0$ with $S^{-1}$ from the left
and insert $1=SS^{-1}$ to obtain
\begin{equation}\label{app8}
 iS^{-1}\gamma^{\mu} S \partial_{\mu}S^{-1} \psi -m S^{-1} \psi =0.
\end{equation}
Using (\ref{app7}) and (\ref{app2}), this can be written as
\begin{equation}\label{app9}
 (i\Gamma^{\mu}\partial_{\mu} -m)\Psi =0 ,
\end{equation}
which is manifestly Lorentz covariant.

Now let me check that $\bar{\Psi}\Psi=\bar{\psi}\psi$. Similarly to (\ref{app6}), one obtains
\begin{eqnarray}\label{app10}
 \bar{\Psi}\Psi & = & \psi^{\dagger} (S^{-1})^{\dagger} \gamma^0 S^{-1}\psi
= \psi^{\dagger} (S^{-1})^{\dagger} S^{\dagger} \gamma^0 \psi
\nonumber \\
& = & \psi^{\dagger} (SS^{-1})^{\dagger} \gamma^0 \psi
= \psi^{\dagger} \gamma^0 \psi = \bar{\psi}\psi.
\end{eqnarray}
 
Finally, let me demonstrate that the scalar nature of $\Psi^{\dagger}\Psi$ 
is not in contradiction with the fact that $\psi^{\dagger}\psi$ is not a scalar. 
This is seen from
\begin{equation}\label{app11}
 \Psi^{\dagger}\Psi = \psi^{\dagger} (S^{-1})^{\dagger} S^{-1}\psi \neq \psi^{\dagger} \psi ,
\end{equation}
which is a consequence of the fact that $S$ is not unitary due to (\ref{inverse}).


\begin{thebibliography}{99}

\bibitem{birdel}
N. D. Birrell and P. C. W. Davies, {\it Quantum Fields in Curved Space}
(Cambridge Press, New York, 1982).
\bibitem{GSW}
M. B. Green, J. H. Schwarz, and E. Witten, {\it Superstring Theory II}
(Cambridge University Press, Cambridge, 1987).
\bibitem{parker}
L. Parker and D. Toms, {\it Quantum Field Theory in Curved Spacetime} 
(Cambridge University Press, Cambridge, 2009).

\bibitem{hill}
E. L. Hill and R. Landshoff, 
``The Dirac electron theory,''
Rev. Mod. Phys. {\bf 10}, 87-132 (1938).

\bibitem{bd1}
J. D. Bjorken and S. D. Drell, {\it Relativistic Quantum Mechanics} 
(McGraw-Hill, New York, 1964).
\bibitem{schweber}
S. S. Schweber, {\it An Introduction to Relativistic Quantum Field Theory}
(Row, Peterson and Company, New York, 1961).

\bibitem{sakurai}
J. J. Sakurai, {\it Advanced Quantum Mechanics} (Addison-Wesley Publishing Company, Massachusetts, 1967).
\bibitem{zuber}
C. Itzykson and J.-B. Zuber, {\it Quantum Field Theory} (McGraw-Hill, New York, 1980).  
\bibitem{zee}
A. Zee, {\it Quantum Field Theory in a Nutshell} (Princeton University Press, Princeton, 2010).

\bibitem{messiah}
A. Messiah, {\it Quantum Mechanics II} (North-Holland Publishing Company, Amsterdam, 1962).
\bibitem{jauch}
J. M. Jauch and F. Rohrlich, {\it The Theory of Photons and Electrons} 
(Springer-Verlag, Berlin, 1976).
\bibitem{greiner}
W. Greiner, {\it Relativistic Quantum Mechanics} (Springer, Berlin, 1990). 

\bibitem{landau}
V. B. Berestetskii, E. M. Lifshitz, and L. P. Pitaevskii, {\it Quantum Electrodynamics}
(Pergamon Press, Oxford, 1982).
\bibitem{ryder}
L. H. Ryder, {\it Quantum Field Theory} (Cambridge University Press, Cambridge, 1985).
\bibitem{weinberg}  
S. Weinberg, {\it The Quantum Theory of Fields I} (Cambridge University Press, 1995).

\bibitem{nikIJMPA10}
H. Nikoli\'c, 
``QFT as pilot-wave theory of particle creation and destruction,''
Int. J. Mod. Phys. A {\bf 25}, 1477-1505 (2010); arXiv:0904.2287.
\bibitem{nikBOOK12}
H. Nikoli\'c, 
``Relativistic quantum mechanics and quantum field theory,''
in 
X. Oriols and J. Mompart (eds.), {\it Applied Bohmian Mechanics: From Nanoscale Systems to 
Cosmology} (Pan Stanford Publishing, 2012); arXiv:1205.1992.
\bibitem{nik_timeprob}
H. Nikoli\'c, 
``Time and probability: from classical mechanics to relativistic Bohmian mechanics,''
arXiv:1309.0400.



\end{thebibliography}
\end{document}